# Transmission Line Pulse (TLP) as integrative method for the investigation of ultra-fast trapping mechanisms on high-k MIM

L. Merlo[a*], I. Rossetto[a], L. Cerati[a], G. Ghidini[a], A. Milani[a], F. Toia[a], R. Piagge[a], L. Di Biccari[a], E. Gevinti[a], G. Croce[a], A. Andreini[a]

[a]STMicroelectronics, SMART POWER Technology R&D, Via C.Olivetti 2, 20864 Agrate Brianza, MB, Italy

**Abstract**

This paper discusses the transmission line pulse (TLP) analysis, generally used for electrostatic discharge (ESD) device characterization, as high potential usable tool also for non-ESD structures.

TLP technique, combined with DC and pulsed I-V characterization, is performed to study the contribution of trap states on current conduction in metal-insulator-metal (MIM) capacitors with an HfAlO stack. The importance of the above mentioned methods is demonstrated by comparing two generations of samples with slightly different charge trapping mechanisms; their impact on the current conduction is furthermore studied by decreasing the pulse width down to 50nsec.

TLP analysis is finally discussed as interesting method to investigate the influence of trap states on the device robustness. The evaluation of breakdown voltage for different time pulses allows to discriminate whether different failure mechanisms occur or not and to establish the impact of trap states on short-term reliability.

## 1. Introduction

IN analog and Smart Power circuits multiple functions, like decoupling, converters or filtering, require the use of capacitors with excellent voltage linearity and matching accuracy, minimizing the size dedicated to them in the IC.

Metal-Insulator-Metal (MIM) capacitor structures are known to be an efficient solution to realize these functions: the challenge is the research and identification of suitable and reliable new materials as insulator layer, in order to maximize the capacitance density.

HfAlO offers interesting performance from this perspective with respect to $SiO_2$ or $Si_3N_4$, consistently to the band-gap and the high-k value reported, [1-3]. Although further materials (such as $Ta_2O_5$ or $La_2O_3/CeO_2$) demonstrate a higher capacitance density [4-7], HfAlO takes advantage of the low leakage current density for the higher band offset [1-3],[8]. This latter is crucial at the high temperature reached in operating condition. The choice of high-k insulating material finally mirrors the necessity of reducing the non-linear behavior (e.g. non linearities in the C-V curve) and frequency dispersion characteristics ascribed to charge trapping phenomenon [3,7]. Accurate tools for investigation and detection of trap states are therefore essential for a continuous improvement of the performance according to the requirements imposed by the final application [9-11].

The Ultra-fast Pulse Measure Unit incorporated into parameter analyzer shows a limited resolution and a not stable pulse in the range of ns due to external set-up (cable, etc.).

Alternatively, due to the well-defined physics of generation and conduction of square pulses, the TLP technique has a high potential usable also for the characterization of non-ESD structures [12-14]. The TLP stress applied to the DUT (Device Under Test) is well repeatable on the same system and reproducible on different systems, delivering more complete results according to an additional lower time stress range (from 1ns to 1.5μs).

The aim of this paper is to discuss TLP analysis as integrative method for the study and detection of trap contribution in MIM structures composed by a high-k insulating layer. The use of combined TLP and pulsed I-V techniques for the evaluation of trapping on capacitors current conduction and reliability is also evaluated.

---

\* Corresponding author. luca.merlo@st.com
Tel: +39 039 603 6157

## 2. Experimental details

### 2.1 Details about tested structures

Devices under test are MIM capacitors composed by a high-k HfAlO dielectric. These structures are characterized by a physical dielectric thickness of 300 Å and an area of $5 \cdot 10^{-4}$ cm$^2$. Two generations of samples, hereafter called Gen I and Gen II, are compared; these devices are nominally identical, except for the different response to charge trapping effects due to different techniques used for the high-k layer deposition.

### 2.2 Information about standard characterization

In addition to the TLP technique, the current-voltage (I-V) behavior is evaluated by means of a Keithley 4200 parameter analyzer. Direct current (DC) analysis uses an integration time of 20 ms while pulsed measurements scan several pulse width values ranging from 1 µs to 20 ms. A duty cycle of 50% and a rise/fall time of 100 ns are considered.

### 2.3 TLP analysis

TLP method combines the plot of the I-V characteristic with a leakage check for each square pulses increasing in step. The characterization technique is based on charging and discharging of a transmission line (coaxial cable) with a certain impedance (generally 50 Ω) depending on material and geometry of the conductors. The inject stress duration in the DUT is determined by the line cable length, while stress rise time is related to parasitic capacitive and inductive contribution of the system [15-16].

Starting from single voltage and current waveforms versus time evaluated on the DUT (Figure 1), the quasi-static values are obtained averaging the waveforms inside a plateau region (highlighted in green), where the signal is expected to be stable [17]. Waveforms are monitored during pulsed and TLP measurements, in order to guarantee that current is integrated at a stable value and that the charge of the capacitor can be considered as negligible.

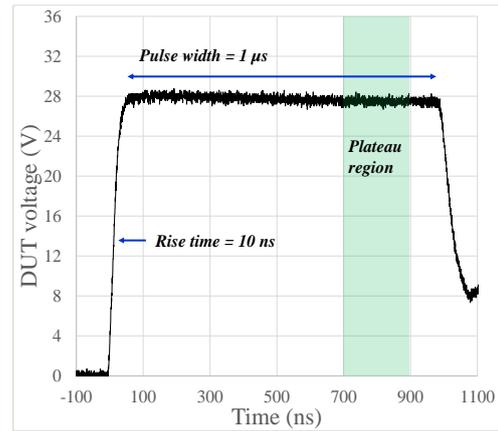

**Figure 1:** TLP transient voltage waveform for a representative case (plateau region highlighted in green).

In this work, transmission line pulser equipment is used to characterize MIM capacitors in order to investigate their conduction and failure mechanisms. These measurements are performed by means of a wafer-level TLP equipment (HPPI TLP-3010C) generating square pulses with fixed rise time (10 ns) and variable duration (50 ns, 100 ns, 500 ns, 1 us). A two probes system (single signal and ground probes) is implemented, where the contribution of parasitic resistances is automatically taken into account by means of calibration routines integrated in the equipment software. After each TLP pulse a leakage current test at 2 V is performed in order to investigate the device integrity; the bias is chosen to monitor the current behavior at low fields. Failure criteria is defined as the increase of the leakage current with respect to the measured initial value and/or the induced snapback effect by the oxide breakdown. Current and voltage waveforms are evaluated by means of a 30 GHz bandwidth scope with a sampling rate of 100 Gsample/sec.

## 3. Study of the conduction mechanisms

### 3.1 Current conduction mechanism

Figure 2 discusses the current-voltage (I-V) behavior measured at room temperature in two representative devices belonging to Gen I and Gen II. Both polarities are discussed.

I-V curves measured at different chuck temperatures (not shown) clarify nature of current conduction mechanisms in Gen I, according to models already described in literature [18-19]. Under positive bias hopping mechanism plays a non-negligible role at low fields (V < 15 V) with an activation energy of 0.18 eV. At intermediate fields (V > 15 V), the current is mainly ruled by Fowler-Nordheim (FN) tunneling. The current variation observed with different chuck temperatures ($T_{chuck}$) evidences that a further mechanism occurs; this latter is thermally activated and presumably ascribed to a trap-assisted phenomenon.

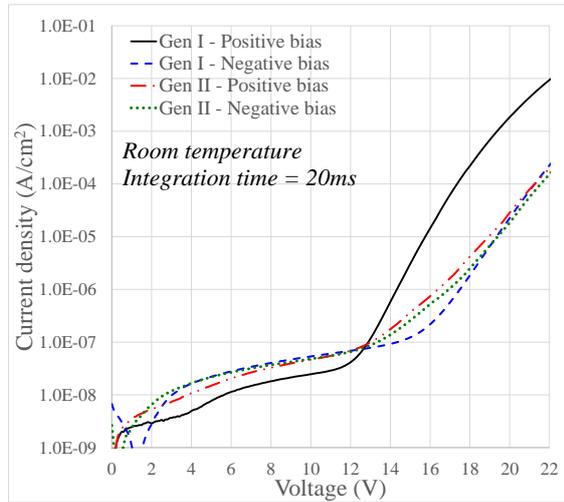

**Figure 2: Current-voltage behavior measured at room temperature in Gen I and Gen II devices under both polarities. The impact of trap states is mainly highlighted under positive bias.**

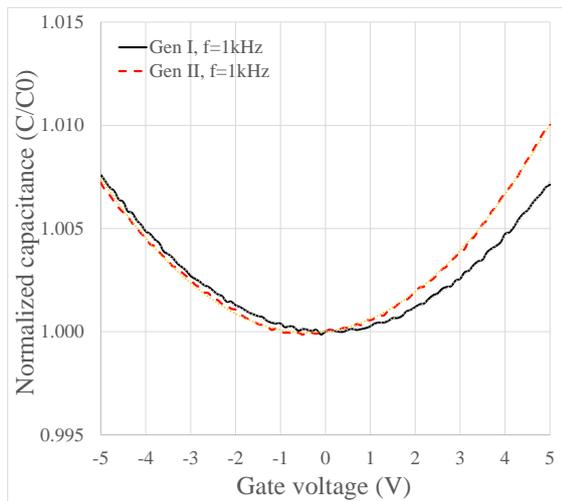

**Figure 3: Capacitance-voltage behavior measured at room temperature in Gen I (black solid line) and Gen II devices (red dashed line) under both polarities.**

Under negative bias the temperature-dependent behavior indicates that a trap-assisted mechanism, presumably ascribed to hopping or space charge limited current, is prevalent. A corresponding trap level of 0.3 eV is calculated.

I-V curves tested in Gen I and Gen II samples indicate that discrepancies are mainly visible under positive bias. The lower current detected in Gen II devices and the different I-V behavior suggests that trapping mechanisms may play a role.

Consistent results are observed in C-V analysis (Figure 3). Gen I and Gen II devices report a different C-V non linearity exclusively under positive bias, thus suggesting that trap states, presumably located in the bottom interface, have a non-negligible contribution.

Although DC and C-V measurements may provide fruitful information about temperature dependence and correspondence of current conduction with previous models described in literature, different techniques are necessary to validate whether different trapping mechanisms occur or not.

### 3.2 Investigation of the impact of trap states by means of TLP and pulsed analysis

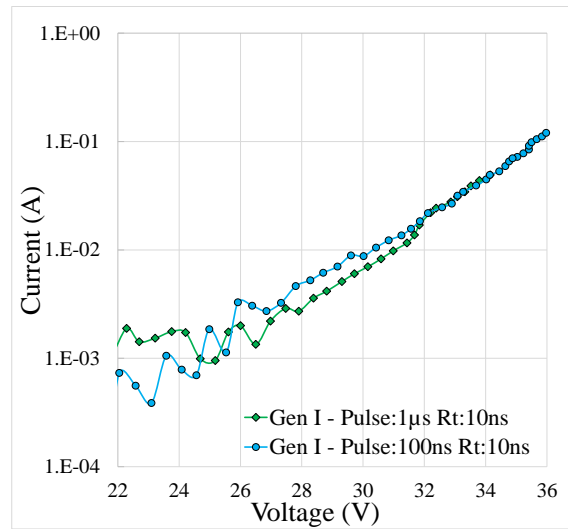

**Figure 4: 100ns-TLP and 1µs-TLP characterization (zoomed view). A representative example for Gen I devices under positive bias is discussed.**

The presence of trapping mechanism was validated by TLP and pulsed I-V analysis. These analyses are focused on the positive polarity where significant difference between Gen I and Gen II is observed.

Figure 4 reports the 100ns-TLP and 1µs-TLP characterizations for Gen I devices. It is clear from these examples that the main TLP drawback is the lack of sensitivity below 1mA.

Several studies reported in literature discuss the use of pulsed bias techniques to determine, in Schottky diodes, the capture/emission kinetics of free carriers due to deep traps [20]. As indicated by D. Pons [20], the pulse varies the space charge region of a diode and induces a different refill of deep traps according to the applied conditions (e.g. pulse width or bias level). The variation of a trap sensitive parameter, e.g. capacitance or current, mirrors the filling/emission of free carriers and allows to describe trap properties.

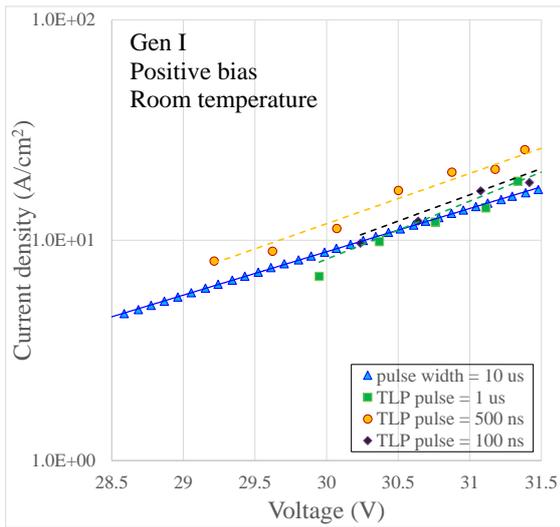

**Figure 5: I-V behavior measured by pulsed and TLP analysis under positive bias for Gen I devices.**

Figure 5 shows a zoomed view of the I-V behavior of Gen I devices under positive bias and measured with different pulse widths below 10 µs. TLP investigation confirms results provided by pulsed I-V analysis, suggesting that trapping plays a non-negligible role in the considered voltage range.

At intermediate fields (figure 6, black squares) the current value increases of about one order of magnitude if we reduce the pulse width from 20 ms to 10 µs. Oppositely, for pulse widths less than 10 µs no further increase of the conduction is observed (figure 5). TLP characterization allows to scan shorter pulse widths down to 100 ns, showing that the conduction characteristics with positive polarity saturates for pulse width of the order of 10 µs and, thus, indicating that the corresponding trap states may have a time constant longer than 10 µs.

Figure 6 compares the correlation between the current density and pulse width (J-pw) measured under positive bias in Gen I and Gen II samples in the time range ≥10 µs.

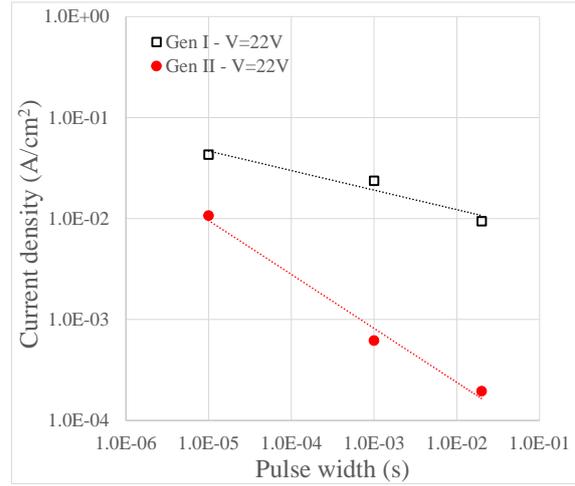

**Figure 6: Correlation between leakage current and pulse width under positive bias.**

The impact of trap states is confirmed by the correlation between current density and pulse width shown in Figure 6. Furthermore, the higher slope of the Gen II devices observed in J-pw curve confirms that this technology is more prone to the effects of trap states, negatively trapped, consistently to the lower current reported in DC analysis by these latter.

The comparison of Figure 2, Figure 5 and Figure 6 shows that TLP combined with pulsed analysis can be a useful tool to analyze different generation of devices and compare the discrepancies induced in these latter by the contribution of trap states, providing further information with respect to DC and C-V analyses.

Similar considerations can be done for I-V curves tested under negative bias (not shown). A good correlation between current density and pulse width is indeed observed, indicating the influence of trap states. Under negative bias Gen I and Gen II demonstrate a comparable current variation with pulse width. For this reason only the analysis under positive bias is discussed.

## 4. Demonstration of failure mechanism time dependence

TLP and pulsed analyses are finally used to discuss breakdown voltage data in Gen I and Gen II samples. Figure 7 summarizes results under positive bias.

Device robustness is defined for several time ranges from 50 ns to 20 ms by using different techniques, highlighted in Figure 7. The breakdown voltage ($V_{BD}$) is defined according to the following failure criteria: a variation of the measured current in more than one order of magnitude for pulsed and dc evaluation and/or the hard leakage increase in TLP analysis.

Gen I and Gen II devices reported an analogous behaviour in terms of breakdown voltage measured by the above mentioned techniques under different time ranges. The device robustness scales linearly with the applied pulse width, indicating that the failure mechanism is time dependent and, thus, that the breakdown voltage scales with the signal frequency. Moreover, the good correlation observed between these two parameters in the whole time range suggests that the same failure mechanism takes place regardless of time range considered.

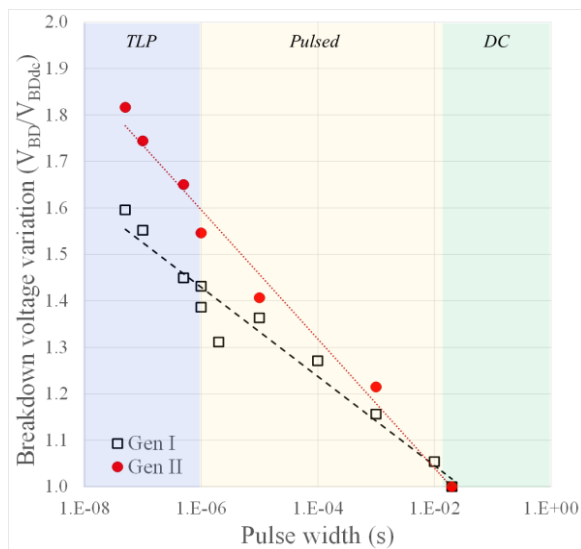

**Figure 7: Variation of the robustness (breakdown voltage) with a different pulse width applied. The behavior measured under positive bias for Gen I (black squares) and Gen II devices (red circles) is shown.**

The use of TLP technique allows to investigate the origin of the failure mechanism down to shorter time (below 50 ns). Moreover, by studying the correlation between $V_{BD}$ and considered time, TLP method allows to study whether different mechanisms occur at shorter times and/or to discriminate whether these latter are influenced by trap states or not.

Figure 7 furthermore compares robustness of Gen I and Gen II devices under positive bias. The more significant increase of the DC normalized breakdown voltage observed when the pulse width is reduced can be attributed to the impact of higher trapping in Gen II compared to Gen I.

## 5. Conclusion

This paper analyses the importance of TLP technique in the evaluation of the electrical performance of MIM high-k with an HfAlO insulating layer. It is presented as integrative tool for the evaluation of current conduction, trap contribution on the electrical performance, breakdown voltage and origin of failure mechanism in short-term reliability.

The use of different measurement techniques is demonstrated to be a useful method to study and analyse the effects of trap states on different generation of devices. Furthermore, the behaviour of the robustness under different time ranges applied allows to discriminate whether different failure mechanisms occur or not and to provide a fruitful investigation on the impact of trapping phenomenon on breakdown voltage.